\documentclass[acmsmall,nonacm]{acmart}
\AtBeginDocument{%
  }

\usepackage{enumitem}
\setlist[itemize]{noitemsep, topsep=0pt}
\usepackage{float}
\usepackage{hyperref}

\renewcommand\footnotetextcopyrightpermission[1]{}

\begin{document}

\title[Unremarkable to Remarkable AI Agent:  Exploring Boundaries of Agent Intervention]{Unremarkable to Remarkable AI Agent: Exploring Boundaries of Agent Intervention for Adults With and Without Cognitive Impairment}

\author{Mai Lee Chang}
\affiliation{%
  \institution{Carnegie Mellon University}
  \streetaddress{5000 Forbes Ave}
  \city{Pittsburgh}
  \state{PA}
  \country{USA}
}
\email{maileec@andrew.cmu.edu}

\author{Samantha Reig}
\affiliation{%
  \institution{University of Massachusetts Lowell}
  \streetaddress{1 University Ave}
  \city{Lowell}
  \state{MA}
  \country{USA}
}
\email{Sam_Reig@uml.edu}

\author{Alicia (Hyun Jin) Lee}
\affiliation{%
  \institution{Carnegie Mellon University}
  \streetaddress{5000 Forbes Ave}
  \city{Pittsburgh}
  \state{PA}
  \country{USA}
}
\email{hlee3@andrew.cmu.edu}

\author{Anna Huang}
\affiliation{%
  \institution{Texas A\&M University}
  \streetaddress{400 Bizzell St}
  \city{College Station}
  \state{TX}
  \country{USA}
}
\email{annahuang5668@tamu.edu}

\author{Hugo Simão}
\affiliation{%
  \institution{Carnegie Mellon University}
  \streetaddress{5000 Forbes Ave}
  \city{Pittsburgh}
  \state{PA}
  \country{USA}}
\email{hsimao@andrew.cmu.edu}

\author{Nara Han}
\affiliation{%
  \institution{Carnegie Mellon University}
  \streetaddress{5000 Forbes Ave}
  \city{Pittsburgh}
  \state{PA}
  \country{USA}}
\email{narah@andrew.cmu.edu}

\author{Neeta M Khanuja}
\affiliation{%
  \institution{Carnegie Mellon University}
  \streetaddress{5000 Forbes Ave}
  \city{Pittsburgh}
  \state{PA}
  \country{USA}
  }
\email{nkhanuja@andrew.cmu.edu}

\author{Abdullah Ubed Mohammad Ali}
\affiliation{%
  \institution{James Madison University}
  \streetaddress{800 S Main St}
  \city{Harrisonburg}
  \state{VA}
  \country{USA}
  }
  \email{mohammau@dukes.jmu.edu}

\author{Rebekah Martinez}
\affiliation{%
  \institution{Portland State University}
  \streetaddress{1825 SW Broadway,}
  \city{Portland}
  \state{OR}
  \country{USA}
  }
\email{becky7@pdx.edu}

\author{John Zimmerman}
\affiliation{%
  \institution{Carnegie Mellon University}
  \streetaddress{5000 Forbes Ave}
  \city{Pittsburgh}
  \state{PA}
  \country{USA}
  }
\email{johnz@andrew.cmu.edu}

\author{Jodi Forlizzi}
\affiliation{%
  \institution{Carnegie Mellon University}
  \streetaddress{5000 Forbes Ave}
  \city{Pittsburgh}
  \state{PA}
  \country{USA}
  }
\email{forlizzi@cs.cmu.edu}

\author{Aaron Steinfeld}
\affiliation{%
  \institution{Carnegie Mellon University}
  \streetaddress{5000 Forbes Ave}
  \city{Pittsburgh}
  \state{PA}
  \country{USA}
  }
\email{steinfeld@cmu.edu}

\renewcommand{\shortauthors}{Mai Lee Chang et al.}

\begin{abstract}
As the population of older adults increases, there is a growing need for support for them to age in place. This is exacerbated by the growing number of individuals struggling with cognitive decline and shrinking number of youth who provide care for them. Artificially intelligent agents could provide cognitive support to older adults experiencing memory problems, and they could help informal caregivers with coordination tasks. To better understand this possible future, we conducted a speed dating with storyboards study to reveal invisible social boundaries that might keep older adults and their caregivers from accepting and using agents. We found that healthy older adults worry that accepting agents into their homes might increase their chances of developing dementia. At the same time, they want immediate access to agents that know them well if they should experience cognitive decline. Older adults in the early stages of cognitive decline expressed a desire for agents that can ease the burden they saw themselves becoming for their caregivers. They also speculated that an agent who really knew them well might be an effective advocate for their needs when they were less able to advocate for themselves. That is, the agent may need to transition from being unremarkable to remarkable. Based on these findings, we present design opportunities and considerations for agents and articulate directions of future research.
\end{abstract}





\maketitle

\section{Introduction}

Advances in healthcare have resulted in a growing number of older adults living longer and experiencing higher rates of cognitive decline as they age. Mild Cognitive Impairment (MCI) is a type of cognitive decline that goes beyond what is expected as part of healthy aging but does not constitute dementia. Many older adults who experience MCI eventually develop dementia. Currently, about 20\% of adults ages 65 and older in the U.S. experience memory challenges (e.g., MCI, dementia)~\cite{2023alzheimer_special}. Approximately 33\% of people with MCI (PWMCI) develop dementia within five years~\cite{2023alzheimer_special}. Early diagnosis and treatment of MCI can delay or decrease the chances of developing dementia. Unfortunately, MCI is massively underdiagnosed, meaning these interventions do not have as much impact as they could~\cite{mattke2023expected}. 

The rising number of older adults coincides with a shrinking number of youth. The U.S. Census Bureau predicts that by 2034 there will be more people over 65 than under 18~\cite{2018census}. The Alzheimer's Association estimates that between 2020 and 2030, the U.S. will need 1.2 million additional direct care workers to support people living with dementia~\cite{2023alzheimer_special}. They note this is a higher demand for workers than any other occupation in the U.S., and they note the turnover rate for direct care workers is extremely high (99\% for nursing assistants). Family members (e.g., spouses, adult children) provide an extraordinary amount of support, taking on the brunt of caregiving work prior to older adults entering high-cost memory care facilities ~\cite{roberto2013trajectories, stoller2008spouses}. These informal caregivers play an important role in the older adults' care team and often face a wide range of coordination challenges depending on their relationship with the older adult and other caregivers, access to resources and support, and physical and mental barriers~\cite{chen2013caring, schurgin2021isolation}. Since people of lower socioeconomic status (SES) experience higher rates of dementia~\cite{2023alzheimer_special}, they will pay an especially high price as the costs associated with direct care will likely rapidly rise due to increasing demand and diminishing supply. 

Researchers have proposed that artificially intelligent (AI) agents could help. Agents might ease some of the caregiving burdens by collaborating and coordinating with informal caregivers over the long-term. Agents could also provide direct cognitive support to older adults in cognitive decline such as providing reminders and monitoring what does and does not get done. The impact of this personalized, collaborative technology has the potential to keep older adults in their homes longer, lowering overall time and costs of memory care by keeping older adults in their homes. One major barrier to this solution is designing agents older adults and their caregivers will accept, adopt, and use in the care network ~\cite{guisado2019factors, wrede2021requirements, hvalivc2022factors}. The Computer-Supported Cooperative Work \& Social Computing (CSCW) literature shows that barriers to acceptance include conflicts of interest between the older adult and their caregivers, a lack of familiarity and comfort with computing technology, a lack of trust (particularly with technology considered "intelligent" systems), and older adults' often unrealistic beliefs about their own capabilities and denial about their own decline~\cite{li2023privacy, caldeira2017senior, soubutts2021aging}.

We wanted to better understand the boundaries that might make agents more or less acceptable to older adults and their caregivers. We refer to boundaries as factors that influence how far the agent will be accepted, such as control, autonomy and trust. These factors could shift agents from being accepted to being rejected by users, and vice versa. For instance, the older adult might accept and trust the agent as long as it operates transparently and respects the older adult's autonomy, but they may withdraw their trust if the agent starts making autonomous decisions without the older adult's explicit consent.  An example is that the older adult wants the agent's support for medical-related tasks but not finances. If the agent takes on finance tasks, that would be crossing a boundary. We also wanted to understand how older adults' and their caregivers' attitudes towards agents cooperating in the care network might change over time. Do older adults who are healthy feel the same way about care providing agents as older adults who are experiencing cognitive decline and must now rely more heavily on informal caregivers? Adopting agents that provide support before memory problems start seems important, in that learning to use new technical systems during a time of cognitive decline seems very difficult~\cite{van2012awareness, guisado2019factors}. In addition, agents could provide better support when they have had time to learn about the people they are supporting---when they understand a family's preferences and patterns of daily living.

We conducted a speed dating study with healthy older adults and with older adults in the early stages of cognitive decline. Declining older adults were accompanied by a primary informal caregiver. The senior centers we recruited participants from were in lower-income areas of the city, and most participants in our study were of lower socioeconomic status. Healthy older adults expressed resistance to adopting an agent before the first signs of any cognitive decline. They worried that adopting an agent would increase their chances of developing MCI or dementia. Paradoxically, they wanted an agent that would have deep knowledge of their preferences and patterns of living if they should start to experience decline. The declining older adults and their caregivers recognized the benefits of having an agent; however, they expressed concerns about its role and its access to personal information. Surprisingly, the older adults in decline suggested that an agent with a deep knowledge of their preferences could advocate for their needs and desires when they could no longer do this for themselves. 

Our paper makes two contributions. First, it reveals new barriers and opportunities around the adoption and use of agents to help older adults \textit{age in place} by collaborating with different stakeholders in the care network. Our work searches for boundaries that demarcate agent acceptance at different points along an older adult's aging journey. We compare the perspectives of healthy older adults to those of declining older adults and their primary informal caregivers. Boundaries of agent intervention are dynamic and influenced by changes in the older adult's abilities, especially cognitive. Second, our paper contributes novel implications for agent design and future research through reflections on the findings. Our findings suggest that agents' roles and capabilities need to evolve based on the dynamics of the boundaries. We hope that these findings contribute to a near future where agents can interact with older adults and their caregivers in fluid, beneficial ways.

\section{Related Work}
In the CSCW, Human-Computer Interaction (HCI), and aging literature, the term “aging in place” refers to older adults who live independently in their own homes for as long as possible, delaying having to move to an assisted living institution or live with a caregiver~\cite{wiles2012meaning, caldeira2017senior, soubutts2021aging}. CSCW and HCI have a history of exploring technology to help older adults age in place including agents that can facilitate collaboration and care coordination within care networks~\cite{barg2017understanding, tiwari2011feasibility, kubota2022cognitively, blythe2010age, wong2024voice, lazar2017critical}. Prior work explores the roles that social robots and agents might play in older adults' lives and how that might impact adoption and continued use~\cite{trajkova2020alexa, blocker2020digital, shibata2011robot, hawkley2019us, gasteiger2021friends}. While most work has focused on healthy older adults, a smaller but growing body of work focuses on older adults experiencing declines in health ~\cite{stogl2019robot, astorga2022social, thoolen2022livingmoments}.  
 Research suggests that agents can help with cognitive decline, as well as coordination of care giving~\cite{zubatiy2021empowering}. Agent support can be categorized as 1) functional or as 2) social and emotional. Agents can detect functional needs, learn about the social situation, and help reinforce what older adults and their caregivers value and feel is important~\cite{houben2019foregrounding, thoolen2022livingmoments, guisado2019factors, wrede2021requirements}. We expand on this work by exploring if older adults in cognitive decline and their informal caregivers might accept help from agents to enhance their independence and overall quality of life.
 
There is a rich literature that focuses on using technology to support older adults~\cite{forlizzi2004assistive}, improve their physical and cognitive function~\cite{mynatt2001developing, barg2017understanding, kubota2022cognitively}, and help reduce the burden on family members. In recent years, research has explored how technology might be used to address interpersonal, social, and emotional needs in complex settings~\cite{shibata2011robot, rogers2017envisioning, gasteiger2021friends, el2020virtual}. In addition to technology support for older adults, researchers have studied ways to help reduce the burden of care for family members~\cite{lindqvist2018contrasting, wrede2021requirements}. Other work explores barriers that slow or prevent agent adoption by older adults~\cite{blocker2020digital, trajkova2020alexa, harris2021smartphone, wrede2021requirements}. Existing work suggests that how agents reason about boundaries may impact adoption~\cite{luria2020social, reig2020not}. Insights from these bodies of literature provide valuable perspectives that motivate our work, ground the scenarios in our study, and help to frame our findings.

\subsection{Technology for Older Adults}
\textbf{Functional support:}
Research has investigated how robots might provide older adults with functional support for their tasks of daily living, such as improving eating~\cite{mccoll2013meal}, exercise~\cite{barg2017understanding}, medication management~\cite{tiwari2011feasibility}, and cognitive stimulation~\cite{kubota2022cognitively}. Existing research explored the use of virtual coaches to support older adults' physical and nutritional health~\cite{siewiorek2012architecture,el2020virtual}. These virtual coaches are in the form of apps, conversational agents, and robots. Smart home technologies such as stair lifts have been shown to support older adults' mobility~\cite{soubutts2021aging} and motion sensors have been shown to provide valuable information about changes in older adults' physical activity~\cite{caldeira2023compare}. Self-tracking technology can also assist older adults with self-care, including monitoring their fitness and health~\cite{caldeira2017senior}.   

\textbf{Social and emotional support:}
Agents and robots can support social and emotional challenges that come with aging~\cite{hawkley2019us, gasteiger2021friends, wong2024voice}. Gasteiger et al.~\cite{gasteiger2021friends} conducted a literature review on social robots and computer agents that combat loneliness in older adults. Previous research showed that virtual coaches can be effective in maintaining or improving older adults' social and emotional health~\cite{siewiorek2012architecture, el2020virtual}.

\textbf{Technology adoption:}
Needfinding studies suggest that older adults generally prefer robot help for physical chores and information gathering, and they prefer human help for personal care and leisure activities~\cite{smarr2012older}. Factors that influence the continued use of agents by older adults include how well the agent performs, the effort to use the agent, pleasure that comes from mastering new technology, and other benefits they experience from agent use~\cite{beer2011mobile, trajkova2020alexa, koon2020perceptions, harris2021smartphone}. Harris et al.~\cite{harris2021smartphone} noted that older adults felt smartphones and conversational agents like Alexa were useful and enjoyable; however, older adults could struggle to become competent users and they often did not understand the privacy risks. 

Other factors impacted caregivers' acceptance of support. Hvalic-Touzery et al.'s~\cite{hvalivc2022factors} literature review looked at factors that influence caregivers' acceptance of assistive telecare systems both before and after deployment. Pre-deployment factors included ethical issues and affordability. After deployment, the important factors included the caregivers' concerns about what the technology could do, the older adults' specific needs, and the overall coordination of care. For adoption of home monitoring systems, conflicts between older adults and their adult children regarding the sharing of the data can pose a challenge to adoption~\cite{li2023privacy}. There are also still open challenges that impact acceptance specifically for older adults experiencing health declines. 

\subsection{Technology for Older Adults With Cognitive Impairment}
PWMCI often struggle to perform the instrumental activities of daily living (IADLs), especially tasks relating to communication (phone calls), transportation (driving), financial management like paying bills, and making and keeping appointments~\cite{mis2019heterogeneity, guo2020instrumental}. This decline creates a lot of work for family members who take on the burden of daily caregiving~\cite{corcoran1994management, roberto2013trajectories}. Treatment for PWMCI focuses on longitudinal monitoring of cognitive and functional status~\cite{gates2013effect, hughes2013engagement}, which can further increase caregiver burden~\cite{corcoran1994management, roberto2013trajectories} .  

\textbf{Functional support:}
One approach that researchers have pursued is designing technologies that aim to delay cognitive decline associated with memory problems through training programs~\cite{pino2020humanoid} and games~\cite{manca2021impact}. One study used a robot to support a psychologist in a memory training program and found that for participants with MCI, the robot led to better performance on visual gaze tasks and stronger reinforced therapeutic behavior than a psychologist alone~\cite{pino2020humanoid}. In another study about cognitive games to slow decline, participants with MCI were more engaged and caregivers found the atmosphere more dynamic and stimulating with a robot than with a tablet~\cite{manca2021impact}. The other avenue is focused on providing support for IADLs without aiming to delay cognitive decline. For example, researchers have used robots to help older adults with memory problems preserve their motor skills~\cite{stogl2019robot} and maintain eating~\cite{astorga2022social}. Mathur et al.~\cite{mathur2022collaborative} designed a routine medication management system for older adults with MCI. Our research falls into this latter category: we aim to better understand what agents can and should be doing to make the lives of people with memory impairment and their loved ones easier.

\textbf{Social and emotional support:}
Researchers have explored unique ways to help older adults with memory problems maintain social connections. For instance, Hoube et al.~\cite{houben2019foregrounding} designed a dementia sound board that uses everyday sounds to trigger past memories and emotional responses to engage with others. \textit{LivingMoments} is a communication system that enables people with dementia to communicate with their relatives~\cite{thoolen2022livingmoments}. Messages from relatives are delivered as printed postcards. Smart speakers have been used to provide reminiscence therapy to individuals with early dementia~\cite{ting2020research}. Kerssens et al.~\cite{kerssens2015personalized} investigated the use of a touch screen computer program called \textit{Companion} to deliver psychosocial interventions to older adults with dementia in the form of images, music, and messages from trusted individuals. Researchers have also explored how people living with dementia and their caregivers are using social media platforms to provide and receive support~\cite{johnson2022s}.

\textbf{Technology adoption:}
A literature review showed that the main factors impacting the adoption of smart health technology for people experiencing dementia and their informal caregivers are: 1) users' attitudes towards technology, 2) ethical issues, 3) technology-related challenges (e.g., design, perceived usefulness), and 4) condition-related challenges (e.g., cognitive decline, condition acceptance)~\cite{guisado2019factors}. For instance, older adults who have memory problems may deny their disabilities and needs, which results in rejection of any kind of help including technology~\cite{van2012awareness}. Wrede et al.~\cite{wrede2021requirements} conducted a study investigating formal and informal caregivers' expectations and needs of unobtrusive in-home monitoring systems. They proposed that acceptable systems must be able to 1) support prevention and proactive measures, 2) prevent information overload, 3) reduce privacy concerns, and 4) reduce ethical concerns.

\subsection{Boundaries: Agents and Interpersonal Relationships}
There is a growing recognition of the need to consider social complexity when designing and evaluating interactions between people and technology. In families, different people have different expectations about how an agent should behave~\cite{luria2020social}. For a conversational agent to be adopted, it needs to successfully navigate issues of who to be accountable to and when, whether or not to keep secrets, and whether or not to ``take sides''~\cite{luria2020social}. It should also be aware of who is present and how that might affect what is or is not appropriate to say~\cite{luria2020social, reig2020not}. Lindqvist et al.~\cite{lindqvist2018contrasting} explored the boundary of when technology is a hindrance versus when it is a solution for people with memory problems from the perspective of professional caregivers. They found that technology is perceived to be a desirable solution for issues around transportation and finance management, but not for social support. Because consequences of memory changes for social interaction and relationships are more pronounced in people with cognitive impairment~\cite{parikh2016impact}, and because having or supporting someone with cognitive impairment involves dealing with sensitive topics, this perspective is especially important in the context of memory decline. Our study addresses this social complexity in that all scenarios are designed to include several dimensions and our participants include both older adults experiencing cognitive impairment and their informal caregivers.

\section{Method}

The goal of our study is to explore and reveal what might be the boundaries of acceptable and unacceptable agent intervention and how these boundaries may change, starting from when older adults are healthy to when they experience cognitive decline. This includes exploring healthy older adults' anxiety that in the near future they might suffer a decline, along with their anticipation of the adjustments needed after the onset of cognitive decline. We followed a research-through-design approach. We chose to conduct a speed dating study using needs validation, which probes desires and values by using storyboards of possible futures~\cite{odom2012fieldwork, zimmerman2017speed}. 

Speed dating as a design research method offers participants many sips of provocative futures using storyboards that illustrate familiar situations with unexpected technology interventions. Following each encounter, the researcher asks questions intended to confirm or reject the validity of the underlying need addressed in the storyboards, reveal how the interactions depicted stack up against norms and expectations, and generate ideas about opportunity areas and risks. This method is often employed in early-stage design explorations~\cite{holstein2019co, luria2020social, reig2021social} where researchers are uncertain about how technology should behave. We spoke to two groups of people, older adults who view themselves as healthy and independent, and a group who saw themselves as currently experiencing cognitive decline. We collected data from 16 healthy older adults and 9 pairs consisting of an older adult with memory problems and their primary caregiver (family member).

\begin{table}[htbp]
	\centering
    \small
	\begin{tabular}{p{0.12\textwidth}  p{0.4\textwidth}  p{0.4\textwidth}} 
 \textbf{Domain} & \textbf{Sample agent behaviors} & \textbf{Sample probing questions} \\ 
    \toprule  
 Managing \newline finances & \begin{itemize}
     \item An agent tells an older adult that she has been giving her niece an unusually large amount of money lately. 
     \item An agent reports an older adult’s overspending to their spouse. 
     \item An agent notices that an older adult has not paid their bills and recommends that they switch to autopay.  
     \end{itemize} & \begin{itemize}
         \item Should an agent monitor for and protect against relatives taking financial advantage of older adults with cognitive decline? 
         \item Should an agent manage families’ budgets and report unusual activity? 
         \item Should an agent access bank accounts and make payments?  \end{itemize}\\ 
\hline
 Cooking & \begin{itemize} \item An agent tells an older adult that he has repeated a step in a recipe. \end{itemize} & \begin{itemize}
     \item Should an agent monitor kitchen behaviors and record recipes and note unusual activity? \end{itemize}\\ 
 \hline
 Transportation & \begin{itemize} 
 \item In a conversation between an older adult and her son about the older adult’s potentially unsafe driving, an agent interjects to note additional incidents of unsafe driving. 
 \item An agent checks the schedules of everyone in an older adult’s network and assigns someone to drive the older adult to their medical appointment. 
 \end{itemize} & 
 \begin{itemize} 
 \item Should an agent decide when it is no longer safe for someone to drive? 
 \item Should an agent mediate in conversations about loss of autonomy? 
 \item Should an agent assign responsibilities to different caregivers based on their schedules? \end{itemize}\\ 
 
 \hline
 Socializing & \begin{itemize}
     \item An agent suggests activities that an older adult and their friends (who have various cognitive and mobility impairments) can all participate in. \end{itemize} & \begin{itemize} 
     \item Should an agent keep track of an older adult’s social circle and interests and make relevant suggestions? \end{itemize} \\ 
 
 \hline
 Physical safety & 
 \begin{itemize} 
 \item When an older adult with cognitive decline and a history of wandering off tries to leave his house, an agent locks the door and alerts the older adult’s wife. 
 \end{itemize} 
 & \begin{itemize}
    \item Should an agent enforce restrictions when a caregiver is not at home? 
    \end{itemize} \\ 
 
 \hline
 Health \& \newline hygiene & 
 \begin{itemize} 
 \item During an especially busy time, an agent reminds an older adult to remind her husband to do his daily exercises. 
 \item An agent tells an older adult’s doctor about the older adult’s recent concerning behaviors. 
 \item When an older adult goes days without bathing, an agent reports this to the older adult’s daughter.
 \end{itemize} 
 & \begin{itemize} 
 \item Should an agent help a caregiver keep track of their responsibilities? 
 \item Should an agent report incidents to medical professionals? 
 \item Should an agent keep track of personal hygiene?
 \end{itemize} 
 \\ 
 \bottomrule
 \end{tabular}
 \caption{In our storyboards, we explored agent/bot behaviors across six domains. We conducted semi-structured interviews, asking questions like those in the third column of the table.}
 \label{tab:scenariotable}
 \end{table}\label{tab:scenariotable}

\subsection{Ideation}
We began by brainstorming scenarios about agents providing cognitive support for instrumental activities of daily living to allow people to age in place. We focused mainly on situations involving coordination among multiple members of a family. Through discussion of literature about support for healthy and independent aging, the challenges of caring for a loved one going through a cognitive decline, and existing AI and smart home products to support tasks of daily living, we identified several dimensions to use in scaffolding our brainstorming. These included: 

\textit{Agent's autonomy}: We considered various degrees of autonomy over events and environments.

\textit{Interpersonal relationships and living situations}: We prioritized scenarios involving different kinds of relationships (e.g., parent and adult child, spouses, friends), living conditions (e.g., living alone and far from family, living with a spouse), and support (e.g., having multiple adult children, having no consistent source of support).

\textit{Older adult's independence and health}: Needing support with aging exists on a spectrum~\cite{roberto2013trajectories}. We discretized this spectrum into categories: 1) healthy and independent older adults with no concerns about cognitive impairment, 2) older adults who have just begun to notice or wonder about cognitive impairment, 3) older adults who have MCI, and 4) older adults who have more progressed forms of cognitive impairment.

\textit{IADLs}: Cooking, cleaning, transportation, managing communication with others, and managing finances~\cite{mis2019heterogeneity}. 

We aimed for our final set of scenarios to cover IADLs where an agent might provide some value. This eliminated situations that require physical support. We viewed agents as mainly providing informational and social support. Our goal was to get good coverage of a large design space along these four dimensions. Our team of researchers met multiple times each week over several weeks to generate as many scenarios as possible. Scenarios explored complex social interactions such as agents that prevaricate, refusing to answer questions from family members about the behavior of an older adult in decline, agents that make various tradeoffs (privacy for safety, independence for safety), and agents that help to facilitate or provide social/emotional support. 

\begin{figure*}
    \centering
    \includegraphics[width=\textwidth]{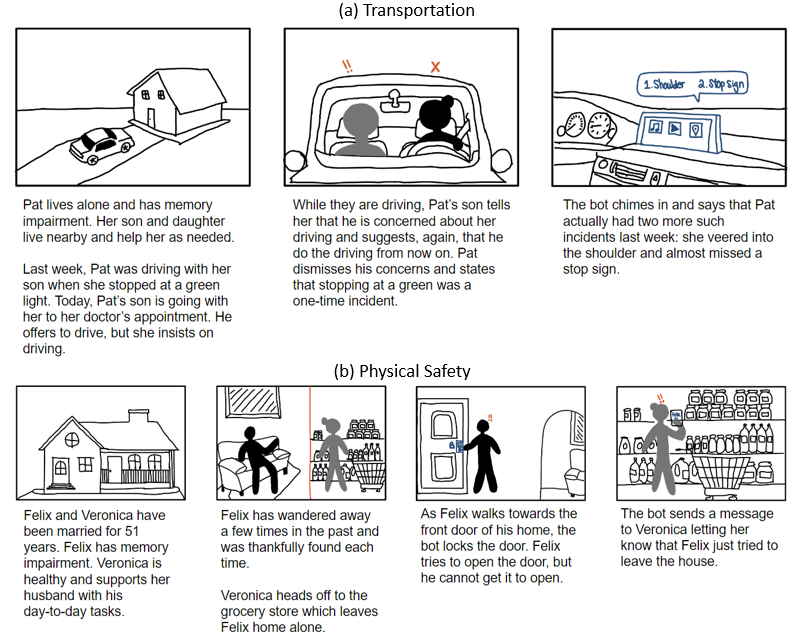}\
    \caption{Two samples from our storyboards. Example (a) depicts a conversation between an older adult and his son about the older adult's potentially unsafe driving and an agent chimes in. Example (b) is about an older adult with cognitive decline and a history of wandering off who tries to leave his house, and an agent locks the door and alerts the older adult's wife.}
    \label{fig:storyboards}
\end{figure*}

\subsection{Final Storyboards}

We filtered a large set of scenarios, synthesizing some and abandoning others. We chose a final set of scenarios to develop into storyboards based on how well they covered and complemented each other along the dimensions described in the previous section and how relatable we believed they were likely to be for our participants. For each storyboard, we generated a lead question that probes the underlying need and intervention to assess if the need is worthwhile or simply a mild annoyance. Table~\ref{tab:scenariotable} contains a summary of the agent behaviors exhibited in our storyboards along with sample questions that we asked participants about each scenario. Figure~\ref{fig:storyboards} shows two samples from our storyboards.

We piloted the storyboards to check that people were reacting to the unmet need and not other aspects of the story such as the technology solution. We did several rounds of piloting. We found that healthy older adults did not relate well to the characters with memory impairment. Since we are interested in understanding their adoption of agent support in their current health state, we modified all the characters to be healthy. In the end, we had two sets of storyboards: one where all the characters are healthy, independent older adults and another where all the characters are experiencing memory impairment. The only difference between these two sets of storyboards was the characters' health. In piloting the storyboards with older adults experiencing memory impairment, we observed that they tend to have a negative bias towards technology described as  ``AI'' and ``agents'', but were more open to the same scenarios when the technology was described as a ``bot''. We therefore revised our storyboards.  The final set of storyboards is in the Appendix.

\subsection{Participants}
We recruited the majority of participants from local senior centers in low SES areas. These centers had low membership fees and were located in economically disadvantaged neighborhoods. We focused on participants with low SES background because they have higher rates of dementia~\cite{2023alzheimer_special}. In addition, low-SES has been overlooked in prior HCI aging in place research~\cite{hu2021toward}. Our recruitment efforts included visiting 40+ local senior centers, and additional participants were recruited by word of mouth. 

\begin{table}[hbt!]
	\centering
    \small
	\begin{tabular}{p{0.05\textwidth}  p{0.05\textwidth}  p{0.05\textwidth} p{0.07\textwidth} p{0.20\textwidth} p{0.15\textwidth} p{0.12\textwidth} p{0.10\textwidth}} 
 \textbf{PID} & \textbf{Pair ID} & \textbf{Age} & \textbf{Gender} & \textbf{Lives with:} & \textbf{Need help with daily cognitive tasks?} & \textbf{Have memory problems?} & \textbf{Support \newline Older Adult?} \\ 

    \toprule

1 & -- & 72 & F & Did not share	& No & No & No \\
2 & -- &	68 & M & Alone & Yes & Yes & Yes \\
3 & -- &	66 & F & Did not share & Yes & Yes & No \\
4 & -- &	72 & M & Did not share & No & Yes & No \\
5 & -- &	73 & F & Did not share	& No & No & Yes \\
6 & -- &	71 & F & Did not share	& No & Yes & No \\
7 & -- &	74 & F & Alone & No & No & No \\
8 & -- &	82 & F & Did not share & Did not share & Yes & No \\
9 & -- &	71 & M & Spouse	& No & Yes & Yes \\
10 & -- & 80 & F & Alone & No & No & No \\
11 & -- & 73 & F & Spouse + Adult Child & No & No & No \\
12 & -- & 79 & F & Alone & No & No & No \\
13 & -- & 80 & F & Alone & Yes & Yes & No \\
14 & -- & 84 & F & Alone & No & No & No \\
15 & -- & 75 & F & Alone & No & No & No \\
16 & -- & 64 & F & Spouse & Yes & No & No \\

 \midrule
 
17 & 1-CG & 64 & F & Spouse +  Adult Child & No & Yes & Yes \\
18 & 1 & 60 & M & Spouse + Adult Child & Yes & Yes* & No \\
19 & 2-CG & 67 & F & Spouse &	No & Yes & Yes \\
20 & 2 & 69 & M & Spouse &	No & Yes & No \\
21 & 3 & 71 & F & Spouse &	Yes & Yes & Yes \\
22 & 3-CG & 72 & M & Spouse &	No & No & Yes \\
23 & 4 & 64 & F & Alone & Yes &	Yes & No \\
24 & 4-CG & 69 & M & Sibling & Yes & Yes & No \\
25 & 5-CG & 80 & F & Spouse & No & Yes & Yes \\
26 & 5 & 83 & M & Spouse & Yes & Yes & No \\
27 & 6-CG & 36 & F & Spouse + \newline Mother-in-law & Yes & No & Yes \\
28 & 6 & 68 & F & Child's Home & Yes & Yes &	No \\
29 & 7-CG & 65 & M & Spouse & No & Yes & Yes \\
30 & 7 & 70 & F & Spouse & No & Yes & Yes \\
31 & 8 & 68 & F & Adult Child &	Yes & Yes & No \\
32 & 8-CG & 20 & M & Mother & No & No &	Yes \\
33 & 9-CG & 76 & F & Spouse &	No & No &	Yes \\
34 & 9 & 80 & M & Spouse &	Yes & Yes & No \\

 \bottomrule
 
 \end{tabular}
 \caption{Participants' demographics: We conducted the study with 34 participants who belong to one of the two groups. The first group was 16 healthy, individual older adults and the second group was 9 pairs consisting of an older adult with memory problems and their caregiver denoted as ``CG''. Group membership was based on their self-perceived health status during the recruitment process. This table shows participants' responses to our demographics questionnaire post recruitment. An asterisk denotes the participant is diagnosed with MCI.}
 \label{tab:participants_table}
 \end{table}\label{tab:participants_table}

We collected data from two groups of participants: 1) older adults who self-identified as being ``healthy'' and 2) older adults who self-identified as experiencing ``memory problems'', accompanied by their primary informal caregiver. The healthy group consists of participants who may still experience cognitive decline, but they view it as typical aging. For the declining group, they view their cognitive decline to be more than typical aging. All of the declining older adults were healthy enough to consent to participate in the study. Table~\ref{tab:participants_table} shows a summary of the participant demographics post recruitment. We collected data from a total of 34 participants who were between the ages of 20 and 84. The majority of participants were low SES and many of them relied on public transit to the senior centers. $78\%$ reported the following types of occupations: community and social services ($12\%$), construction ($6\%$), education instruction and library ($12\%$), food prep and serving ($9\%$), management ($6\%$), office and administrative support ($12\%$), personal care and service ($12\%$), and sales and related occupation ($9\%$). 50\% of participants retired from these occupations. 23 participants identified as female and 11 identified as male. 16 participants were healthy older adults, nine were older adults who have memory problems and nine were caregivers of the older adults with memory problems. Caregivers were asked to focus on the person they support. Overall, we conducted 16 sessions with healthy older adults and 9 sessions with pairs of older adults with memory problems and their caregivers.

\subsection{Procedure}
Study sessions took place in a private room at the senior center or over video chat via Zoom. After obtaining informed consent, a researcher explained the study's goals and procedures. They also defined ``bots'' (using familiar terms like ``Siri'', ``Alexa'', ``robots''). Then, the researcher administered a demographics questionnaire that includes establishing context about the participant's cognitive health, independence, and living conditions. The researcher then showed the participant one storyboard at a time and read the text out loud. Immediately after showing a storyboard, the researcher asked the leading question which frames the discussion around the underlying need. The researcher then asked a few semi-structured interview questions to probe whether our assumptions about agent support align with people's needs and expectations, and target new areas of opportunity and potential points of concern. Based on the progress of the conversation, the researcher selects the next storyboard. Study sessions were audio recorded. Each study session lasted approximately 60 minutes. For each session with healthy older adults, we covered between 2 and 8 storyboards ($M=5.9$). In two sessions, the participants had to leave early and only reviewed two storyboards. For each session with declining older adults and their caregivers, we covered between 2 and 6 storyboards ($M=3.5$). Each participant was paid \$20.00. Our study was approved by an Institutional Review Board.

\subsection{Analysis}
We transcribed all sessions and analyzed responses using affinity diagramming~\cite{beyer1999contextual}. Researchers regularly use affinity diagrams to synthesize data from speed dating studies~\cite{odom2012fieldwork,luria2019re, luria2020social}. Eight researchers met multiple times as a group to collectively arrange the participants’ quotes based on emerging affinity to one another until we reached a consensus on the arrangements. We then reexamined the clusters that emerged based on our research questions to reveal salient themes.

\section{Results}
The majority of our participants were women living with their spouses. Some participants had children and grandchildren who live nearby and other participants' children are scattered in different parts of the U.S. Some participants had impairments to vision, hearing, and mobility in addition to cognitive impairment. Almost all of the older adults took multiple medications. Some of them were caregivers to their parents who suffered from dementia. Most of the participants regularly engaged in routine social activities each week including exercise, Bingo, classes for seniors, Bible study, volunteering, and interactions with family and friends. Almost all have smartphones and use them for communication, navigation, scheduling, photos, information searches, social media, bill payments, and more. Some of them still drive, some never drove, and others used to drive but now take public transportation. Participants who are caregivers frequently spoke of the frustrations of caregiving, including managing schedules, medications, and finances. 

Our notation for quotes is as follows: for older adults, each quote is denoted by their perception of their health and participant number (``H~--~\#'' for healthy, ``D~--~\#'' for declining) and their gender and age (``M'' for male, ``F'' for female). For caregivers, each quote includes their role, participant number, gender, and age (e.g., Supporting~--~Spouse~--~17 (F~--~64)). Our analysis revealed  insights around three primary themes: 1) adoption, 2) fears, and 3) boundary of control and unremarkableness.  

\subsection{Adoption}
Several healthy older adults had a strong, negative reaction to adopting agent support whereas declining older adults and their caregivers were willing to adopt. Many healthy older adults worried adoption of technology meant to support them should they experience cognitive impairment would increase their chances of developing MCI or dementia. For example, they worried that if they got a fall detector, then they would be more likely to have a fall. Declining older adults recognized that the agent could bring awareness to their health decline. Caregivers saw the potential for agent support to provide them quality time with the older adults. 

\subsubsection{Healthy Older Adults' Resistance to Adoption}
Many healthy older adults described why they do not want an agent's support for their current life. Adopting the agent was viewed as ceding control. For example, H~--~7 (F~--~74) explained that she would not want an agent to help with finances: \textit{``Not right now, I think, cause I wanna be able to look at it and I check my bank account almost every day.''} Several healthy older adults expressed that they are hesitant to adopt agent support because they believe that they will never need support. This denial is reflected in the stories that participants shared that revealed a mismatch between their true ability and perceived ability. In one storyboard, an agent notices that an older adult has not paid their bills and recommends that they switch to autopay. H~--~9 (M~--~71) commented, \textit{``You could forget sometimes, but I don't forget, I don't do that''}.   

This mismatch was especially prevalent in the driving domain. We heard stories of how families would pretend things were okay, leaving it to others outside of the family to set limits based on declining capabilities. When discussing if there is a need to keep older adults from driving when they are no longer safe on the road and whose responsibility it is to decide when someone is no longer safe on the road, H~--~11 (F~--~73) shared, \textit{``My mother-in-law, she's 92, and the only reason she's not driving is because they wouldn't do her inspection on her car. But she was at a point where she had knocked off about four side view mirrors and she had scratches on the back of her car, but she wouldn't stop driving. And I don't know why her children didn't say something to her. Well, she probably didn't listen, but she really needed somebody. So the mechanic was the one that really got her to stop. He told her he couldn't inspect her car anymore.''} When asked how an agent may provide support in this situation, H~--~11 (F~--~73) said that her mother-in-law might listen to an agent more than to her family about when to give up driving because the agent is,  \textit{``like a stranger''}, which would allow the mother-in-law to not feel embarrassed and put her pride aside. H~--~11 (F~--~73)'s example also shows that family members seem to be timid at setting limits. Sometimes they work with third parties to create and enforce rules for older adults who are declining.

Most healthy older adults believed that they will be able to recognize their mismatched capabilities and make a good judgment call by themselves, and their family will easily intervene. Regarding making the decision to stop driving, H~--~11 (F~--~73) stated, \textit{``I don't know [about needing agent support now]. I'm pretty observant… my children or my husband would say something about it too… but I think I'll be able to [decide] on my own.''} H~--~11 (F~--~73)'s comment captures how some healthy older adults view themselves as exceptions to the possibility that judgment about one's own capabilities often declines with age.

When asked to reflect on how they feel about the possibility of eventually coming to need assistance with IADLs, participants imagine that an assistive agent will have intimate knowledge about them as if it has been with them over a long period of time. In one storyboard, an agent reported an older adult's overspending to their spouse.  H~--~9 (M~--~71) shared (``he'' refers to the bot in the story):\textit{``He's supposed to [report the overspending], you know, cause [the agent and the older adult couple are] together and they together that long… So he did the right thing, because it could have wound up worse. You know, he could've spent all the money.''}

\subsubsection{Declining Older Adults' and Their Caregivers' Willingness to Adopt}
In contrast, older adults experiencing memory problems and their caregivers were more likely to accept the agent now, and they expressed a desire to adopt it before the start of cognitive decline. When asked for a time estimate of when older adults with memory impairment and their caregivers would want to adopt, Supporting Spouse~--~22 (M~--~72) responded, \textit{``I would say no later than in stage two [basic forgetfulness]… I would say early stage [noticeable memory difficulties] at the latest.''} D~--~21 (F~--~71) who is being taken care of by Supporting Spouse~--~22 (M~--~72) added, \textit{``I think it would be wise to do [adopt bot prior to memory decline].''} In addition, many older adults experiencing memory problems realized that they may not notice their health decline and would want the agent to notify them and their caregivers. In one storyboard, an older adult makes unusual purchases that result in going over their monthly budget; then, the agent detects this and alerts the older adult's child. In response to this, D~--~20 (M~--~69) shared, \textit{``If I reach the point where I’m making mistakes like this frequently, my children need to know.''} D~--~25 (F~--~64) said, \textit{``I wouldn't mind [agent’s support] because I've ran my coffee machine and forgot to put the coffee pot under there and the coffee just dripped out… I make mistakes at my age. I wouldn't have a problem with that [agent]; I wouldn't mind it [agent] chime [in to let me know of my mistake].''}

One of the caregivers, Supporting~--~Spouse~--~33 (F~--~76), explained, \textit{``Life is becoming so complicated that I'm now at the point where I would like to offload some of this stuff onto a helper that we could afford. We can't afford… to hire people to do this stuff for us. If we could find an affordable, reliable system to help us, then we could go to the movies or something like that. Which is what, at this point in our life with few years left, we'd like to have something left to enjoy life and not have to spend so much of our time on these manual ways of handling life.''} Similar to Supporting~--~Spouse~--~33 (F~--~76), several caregivers view the agent as being able to enhance their quality of life by also giving them the gifts of time and attention. They also point to the importance of agent support being affordable.

\subsubsection{Service Gaps as Pathways to Adoption}
Participants in both groups described functional and social gaps in caregiving services that could serve as pathways to adoption. Regarding functional gaps, both groups described agent support for transportation, reminders, fall alerts, and finances. For arranging rides, H~--~13 (F~--~80) expressed that \textit{``Uber is for the rich''}, which reflects that they desire more affordable options than currently exist. Older adults shared that they desire more intelligent fall detectors as the current ones require them to push a button and are prone to false alerts: \textit{``I had the thing [fall detector] around my neck and I'm sleeping in it, and if I hit it [by mistake]… they [family] would call me.''}~-~H~--~4 (M~--~72).

Both groups expressed the need for help with financial management, but they spoke of this in different ways. Healthy older adults discussed their fear of overspending. Some described the agent managing their budget, whereas others preferred that the agent not be involved in finances. They perceive the agent to be an \textit{``outsider''} and believe that it should not be brought into finances because this is a family issue. \textit{``The agent has nothing to do with that [finances]. He [agent] needs to stay out of that, because you are getting into an area where you're dealing with somebody's finances and that can backfire on you.''}~-~H~--~2 (M~--~68). Healthy older adults expressed discomfort with the idea of an agent being the first to contact their friends to ask for repayment of their money. \textit{``I would've probably talked to the friend… and then if they didn't pay, I would've gone to the agent.''}~-~H~--~14 (F~--~84). This suggests that while agents could be helpful by filling some service gaps in finance management, this is also an area in which they will need to be particularly cognizant of boundaries around privacy and relationships.

Social and emotional gaps include memory training coaching and emotional support. H~--~2 (M~--~68) suggested, \textit{``When you build an AI, you have to build an AI to be able to know, to be able to sense depression, to sense stress… you know, to be there.''} Declining older adults also expressed a desire for the agent to detect depression and be a memory coach. \textit{``I have another option that would be for wellness on the bot app, which would be, so [participant's name], how's your memory IQ going today? Have you had any issues today? Like a daily bot, memory checkup.''}~-~D~--~25 (F~--~64). While participants' specific preferences for how agents should function in these services might be different if they were healthy vs. if they were declining, the service domains themselves are largely the same. Therefore, these services can serve as gateways through which agents can be adopted.

\subsection{Fears}
One of the main themes that emerged across many participants is regarding fears, which can shape the boundaries around agent adoption. For example, fear about the future may affect the barriers around when to adopt. We observed differences between the two groups' fears. {Several healthy older adults expressed worries about being a burden to their family and friends and how others might view them. Declining older adults and their caregivers raised concerns about information sharing and privacy. In addition, caregivers conveyed concerns about their caregiving performance.

\subsubsection{Healthy Older Adults' Fear of Being a Burden}
Many healthy older adults' responses hinted that their main fear is being a burden to family and friends. This was reflected in comments about a safe place to live, need for support, and reliance on others. For instance, H~--~9 (M~--~71) commented, \textit{``I know a lot of people that didn't [pay their rent] and got evicted… First thing you do, you gotta have a place to stay. First thing you do is pay your rent every month. You ain't gotta worry about all that. At least you always have somewhere to stay.''} H~--~9 (M~--~71) added that children may be able to help with some things but they cannot take care of everything. Moreover, H~--~4 (M~--~72) described her priorities: \textit{``I always pay my bills… I don't wanna be homeless out there, you know, I'm too old now.''} H~--~4 (M~--~72)'s comment captures how several healthy older adults view bill payments and financial management as key factors to living independently in their homes. Many healthy older adult's also brought up not wanting to burden their family such as H~--~2 (M~--~68) stating, \textit{``Sometimes I don't wanna call my son or other people because I don't want to feel inadequate like I can't help myself.''}  Some of the stories they shared hinted that they would work to make sure others remained unaware of a potential problem and wait for the last minute to get help. A story shared by H}~--~11 (F~--~73) illustrates this common behavior pattern:
\textit{``Because with my knee, I limped for a long time, but I didn't complain. It was hard to get in the tub and stuff like that. And it was at the very last minute when I could hear the bones rubbing. Then I went to the doctor and had my knee done.''} 

\subsubsection{Fear of a Negative Image of the Older Adult}
Several healthy older adults spoke about their self performance and the value that comes from appearing healthy and mentally ``with it''. Some participants' stories suggested that they have fears around ageism. They discussed their lack of familiarity with using technology such as needing to upgrade their phones. In addition, caregivers described protecting the image of the older adult they cared for. They indicated that an acceptable agent needs to hide the older adult's declining condition from other family members and friends. D~--~22 (M~--~69) shared about when he was a caregiver for his father: \textit{``Just the fact that [my father] is not able to manage his finances anymore is a big hit to his self deeply. And again, I wouldn't want the bot to tell everybody [except me].''} Caregivers expressed the desire to control whom the agent shares information with to protect the declining older adult's self image.

\subsubsection{Declining Older Adults' and Caregivers' Fear of the Agent Revealing Their Secrets}
Older adults with memory impairment and their caregivers envisioned the agent being with them over a long period of time. They described the agent as a repository of important information on them that presents both a benefit and risk. They voiced concerns that the agent might report information about their decline, particularly driving, to their caregivers. \textit{``[The bot could] rat you out… they [older adult's children] can go to the bot [and ask,] well, how's her driving been for the last weekend?''} -D~--~21 (F~--~71). Several caregivers worried that the agent may tattle on them to law enforcement. \textit{``Say if you say something very incriminating, [the bot] would just pick up on it and just tell the police''}~-~Supporting Child-36 (M~--~20). They described not wanting to have to monitor their own speech when in proximity to an agent. They expressed a desire for agents to keep secrets from outsiders.

\subsubsection{Caregivers' Fear of Making Mistakes in Caregiving Due to Information Overload}  

Caregivers conveyed worry about their performance in caregiving. They expressed the need to ensure that they are not missing something important. However, they also do not want to be overwhelmed by a constant stream of alerts. They described wanting agents that communicate the right amount of information at the right time to them. Several caregivers expressed concerns about the older adult being in dangerous situations such as unsafe driving that poses harm to themselves and others on the road. \textit{``I think it [alerts from bot] would just stress me to be honest with you, but I wouldn’t want anything to happen to [participant’s husband] either… I woudn't want him to hurt anybody else [unsafe driving]… I mean it’s a catch-22.''}~-~Supporting~--~Spouse~--~17 (F~--~64).

\subsubsection{Differences in Fear of Privacy Breach}
Both groups of older adults raised concerns regarding the possibility of the agent's information getting compromised. However, for the caregivers, they shared that the need for agent support outweighed their fear of having their privacy exposed. When asked to consider the situation involving an older adult who goes for days without bathing and the agent that reports this to the older adult’s daughter, H~--~12 (F~--~79) commented, \textit{``When they say `smart home', it makes me cringe. If I have a smart home, somebody else is gonna be smart enough to get in there.''} In contrast, one of the caregivers, Supporting Children-36 (M~--~20), seemed accepting of this situation, sharing, \textit{``That is kind of a breach, but like it's okay because it [bot] can help you.''} 

\subsection{Boundary of Control and Unremarkableness}
Several storyboards triggered participants to reflect on their desire to maintain control over their own lives and futures and the agent's presence in the care network. Participants described control to involve who controls the agent and who can make decisions that impact the older adult. Control of the agent may also impact the agent's presence and its ability to collaborate and coordinate with stakeholders in providing care. Agents can do work in the background or they can function in ways that make it more obvious what they are doing and how they are providing support to users. This type of agent presence matters and has been referred to as unremarkable computing~\cite{tolmie2002unremarkable}.

We observed that participants' desire for control shifts according to the their perception of their health. Healthy older adults expressed a preference for controlling the agent, while those in decline preferred the agent to assert control on their behalf to advocate for them. Healthy older adults wanted the agent to be unremarkable, whereas declining older adults desired a remarkable agent. Caregivers, feeling responsible for the well-being of the older adult, expressed a desire for control over the agent. 

\subsubsection{Healthy Older Adults Desire to Control Agent}
Healthy older adults expressed the desire to control the agent to operate within a certain space in their care network. They envision agents that take passive roles, only doing what the agent is told to do, never in charge of the older adult: \textit{``The agent [is] supposed to know what's going on with you anyway… [the agent] can only do what you tell 'em''}~-~H~--~9 (M~--~71). \textit{``This thing [agent] is objective, it's simply telling you stuff that happened, as opposed to a son who might be trying to take away your freedom.''}~-~H~--~10 (F~--~80). Healthy older adults viewed the agent's passive role in care coordination as bridging and enhancing the older adults' autonomy and decision-making. Participants also perceived the agent's objectivity as affording more control to them.

\subsubsection{Healthy Older Adults Desire Agent to be Unremarkable}
Most healthy older adults described how the agent could work in the \textit{background} and be unremarkable in supporting their care needs. They view the agent as a safeguard running in the background to prevent errors and automate mundane, menial tasks. For example, H~--~16 (F~--~64) explained, \textit{``When you get to be a certain age, you need that backup [bot]. I mean, you need that… There's nothing wrong with having a second plan… I think that's very important as you get older, that somebody's checking on you. And that's to me like another check.''}~-~H~--~16 (F~--~64). 

\subsubsection{Declining Older Adults Desire Agent to Assert Control on Their Behalf}
Older adults with memory problems described their desire for the agent to take an active role in helping them maintain control over their own lives and futures while cooperating with their caregivers. One of the most unexpected insight is that they want it to advocate for them as their own ability to advocate for themselves begins to diminish. They described the agent collaborating with their caregivers such as communicating and mediating the appropriate information with various stakeholders in the network to provide the best care for the older adult. D~--~30 (F~--~70) stated, \textit{``The bot's basically providing the data for the financial advisor, the attorney, the health professionals to make the needed decisions on part of me, since I am not there mentally anymore. That bot basically is taking care of me and coordinating with my advisors.''}

Older adults with memory problems drew on the storyboards to envision the agent being documented in their legal paperwork that gives the agent permission to actively take part in critical decisions (finances, health, last days) relating to the older adult. This suggests that these older adults fear that the people making decisions for them might not always know their desires or might not always act in accordance with their desires.
\textit{``But all those decisions [regarding critical support for the older adult's health decline] would have to be made while you have your mental faculties, right? Like when you make your will, you can't make a will if you're mentally impaired. An attorney will not let you. Right?… So the same thing for this [bot]. I can see where the bot should be part of your legal paperwork that dictates what you want done, how you want to live your last days, that bot somehow be included legally.''}~-~D~--~30 (F~--~70).

Older adults expressed a desire for the agent to advocate for them by addressing changing power dynamics that can disrupt relationships. H~--~5 (M~--~73) spoke with frustration and concern for a close friend experiencing memory problems. H~--~5 (M~--~73)'s friend feels she was being mistreated because her son and lawyer made changes to her will without any input from her. H~--~5 (M~--~73) commented, \textit{``She [participant’s close friend] feels like she's in a position where she doesn't wanna alienate her son [by finding a new lawyer]… But she feels like, you know, then he won't love me.''} 

H~--~5 (M~--~73) reflected on how an agent might help to avoid or mediate conflict that can arise from a situation like this. She described that if another person becomes involved in the situation, her friend's children will \textit{``get kind of outta whack''}. However, if an agent gets involved, H~--~5 (M~--~73) anticipates that the children \textit{``can't say, oh well her boyfriend's trying to do it to me''}. Most participants suspected that it would be harder to attribute personal intent, desire, or ulterior motives to an agent. They assumed an agent could advocate for the older adult without the risk of negatively impacting a fragile relationship. In the example with the mom and son above, if the agent advocates for the mom, the son could get mad at the agent but the mom does not have to worry about her son not loving her anymore.

\subsubsection{Declining Older Adults Desire Agent to be Remarkable}
Most older adults with memory impairment described the agent working in the \textit{foreground} and being remarkable, taking an active role in the decision-making process pertaining to the older adult's health. In most cases, they envision the agent being a part of their family's conversations to plan for changes in the agent's support. D~--~20 (M~--~69) illustrates this desire: \textit{``So I can envision two discussions. First, the purpose of which is to deal with the family dynamics in a very personal incentive way, then a second meeting in which the bot is there. The family’s already got a pretty good grip of where they think they want to be and then they can present a more coherent and unified set of instructions to the bot… I suppose there's going to be some exchange back and forth between the family and the bot… It's not just, here's what we've decided now, adjust your [bot] programs accordingly.''}

\subsubsection{Caregivers Desire to Control Agent}
Caregivers expressed the desire to control the agent because they felt ultimately responsible for the older adult. This insight aligns with prior research showing that formal and informal caregivers of older adults experiencing dementia want control of the data from in-home monitoring technologies~\cite{wrede2021requirements}. \textit{``I would like to have a common-sense conversation with the bot. But still, at the end of the day, I would like to make my own choice.''} – Supporting~--~Child~--~32 (M~--~20). Supporting~--~Child~--~32 (M~--~20)'s perspective captures several caregivers' preference to be the final decision maker and for agents to not make decisions, especially when the decisions impacted the caregiver. Other caregivers described high expectations around the agent's reasoning and level of participation: \textit{``This is an intelligent bot, and it should be able to evaluate and be able to make decisions or at least recommendations on the progress of that person and whether it needs to be increased or decreased based on interactions with that person. So I would say if it's a good artificial intelligence, they should be a part of the decision-making process as well.''}~-~Supporting~--~Spouse~--~29 (M~--~65).

\section{Discussion}
\begin{figure*}
    \centering
    \includegraphics[width=0.8\textwidth]{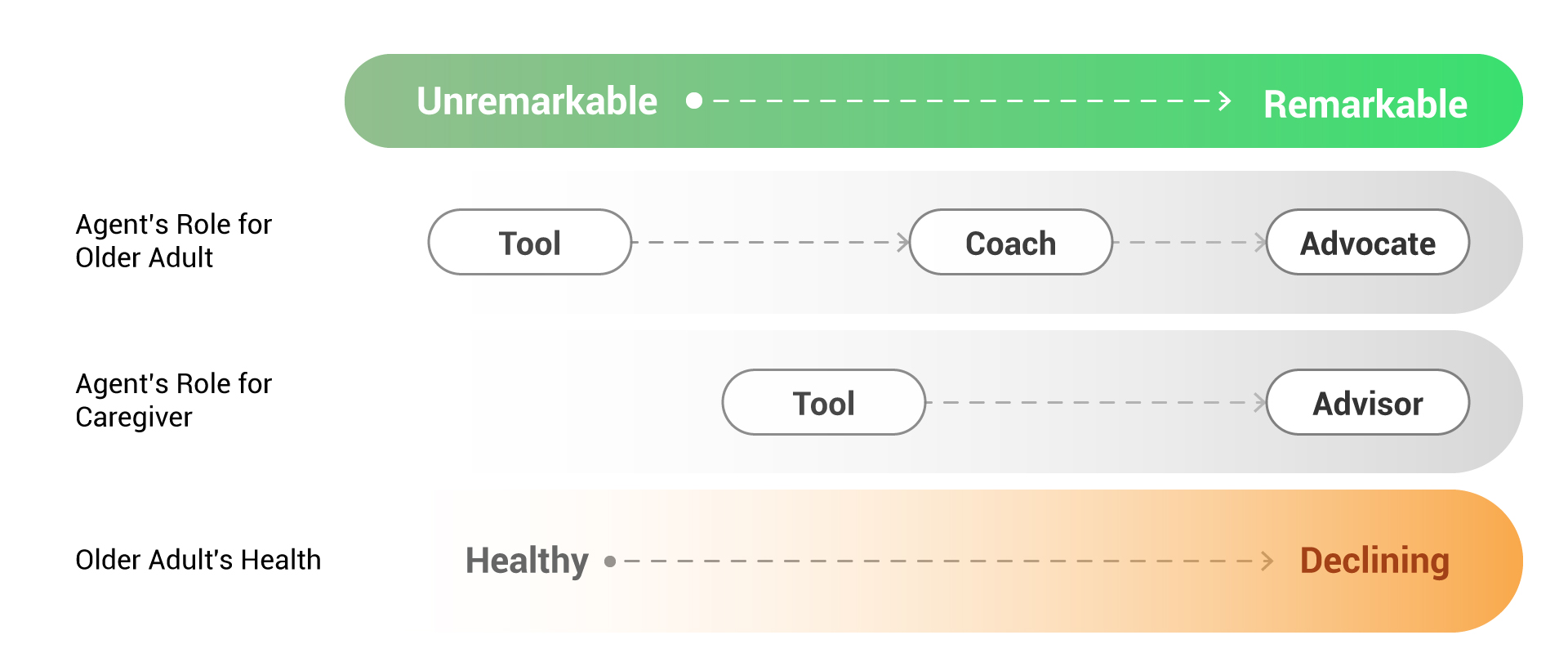}\
    \caption{Summary of design considerations for agent to support care coordination based on our findings. When the older adult is healthy, the agent is unremarkable and serves as a tool. As the older adult's health declines, the agent transitions to be remarkable. For the older adult, the agent shifts to serve as a coach and advocate. For the caregiver, the agent transitions from a tool to an advisor.}
    \label{fig:design_considerations}
\end{figure*}

We explored the boundaries of agent intervention at different life stages of older adults. Our findings show that these boundaries shift according to older adults' stages of decline along with emotions such as fear, anxiety, and uncertainty. Prior work shows that the main factors that influence older adults' tech adoption are related to functional support~\cite{koon2020perceptions, harris2021smartphone, lindqvist2018contrasting}, privacy~\cite{li2023privacy, wong2024voice}, and independence~\cite{caldeira2017senior, soubutts2021aging}. Our work confirms these findings, but also suggests that participants desire agents to \textit{evolve} over time to take more active roles in care coordination. This presents a novel lens for CSCW to use in approaching the design space of agents as facilitators of collaboration. Agents will need to learn to adapt to their users' changing needs including possessing social intelligence to understand the social dynamics and interpersonal relationships within the family. This may be because technology, particularly chat-based systems, has evolved such that people assume agents have the common sense and intelligence to take on these complex and nuanced tasks. Therefore, they may be seen as suitable for addressing the increased need for companionship as people age~\cite{ryu2020simple,valtolina2021charlie}. Declining older adults and their caregivers seemed to have high expectations of what agents might actually do well in the future. Based on our insights, we share design opportunities and considerations and open research questions regarding adoption, the agent's role, the agent's unremarkableness, and people's perception of control. We note that each of these insights pertains to the notion of an agent that evolves over time---in types of functional and social support, social role, and control---to meet its user's (and their loved ones') evolving needs. Figure~\ref{fig:design_considerations} shows a summary of the design opportunities and considerations.

\subsection{Paradox of Adoption}
Healthy older adults expressed resistance to adopting agent support, while older adults with memory problems and their caregivers thought about using it now and using it in ways that require a long-term understanding of routines and preferences. This is a paradox: for the agent to be able to personalize to the declining older adults based on long-term history and understanding, it will need to be adopted before the decline occurs. Healthy older adults' resistance to adoption was likely related to dreading the bleak possibility of cognitive impairment. Denial is commonly used as a defense mechanism to keep a future they do not want at bay. This may be similar to older adults who resist making a will; this is something firms specializing in elder law services frequently discuss in their FAQ targeting adult children~\cite{elder_law}. CSCW research has highlighted tensions around aging and youth with regard to older adults and their caregivers' attitudes toward technology adoption.  When older adults deny functional support, it can result in considerable emotional tension between them and their caregivers~\cite{soubutts2021aging}. Older adults resist adopting self-tracking tools to monitor health activity (exercise) and indicators (blood pressure) because they perceive them to be associated with young people and not designed for them~\cite{caldeira2017senior}. These forms of resistance and tension pose challenges for the ability of assistive technologies to successfully facilitate collaboration and coordination in service of functional tasks, as they are designed to do. Our findings add to this literature by revealing the paradox around intention to adopt future agent technologies, which likely further complicates already-complicated interpersonal dynamics that make collaboration- and coordination- supporting technologies for older adults difficult to design.

\subsubsection{When is the Right Time to Adopt?}
There is an opportunity to rethink adoption in terms of older adults' health stages. Our findings along with prior work~\cite{van2012awareness, guisado2019factors} show the phases of denial: 1) healthy older adults are in denial and may use the agent later, 2) older adults with early memory problems are accepting of the agent and 3) older adults with dementia are in denial and will not adopt~\cite{van2012awareness, guisado2019factors}. Healthy older adults resist adopting agents designed to support them when they do not need support, but want agents to be there for them when they do need it. Therefore, in order for agents to support older adults with memory problems and their caregivers, they will most likely need to be adopted \textit{before} their users think of themselves as ``older adults''. We therefore suggest rethinking how and when we, as researchers, become concerned with adoption, and reorient toward understanding how an  agent can be adopted ``ahead of time'' such that it is already in the older adult’s life by the time they need it. For this approach to work, an agent will need to be initially useful for practical tasks such as scheduling reminders, coordinating transportation, and paying bills, so that it is already familiar to its user by the time it needs to take on tasks of assistance and care. Adoption may need to occur much earlier. It may be beneficial to introduce agents at other times of transition, such as when children start school and their parents begin to encounter complex family logistics around maintaining everyone's increasingly busy schedules---a task for which parents desire technology that can help~\cite{davidoff2010routine}. Later, if someone in the household starts to experience cognitive decline, the agent can transition to handling more complex social tasks, including advocating for the older adult. How it might transition between these roles remains an open question.  

Participants in our study expected a cognitive support agent to provide intimate, personal assistance. There is an opportunity to develop agents that could be ``pre-positioned'' to jump into action when their users are ready---that is, to learn in the background while the older adult is healthy, but wait to take action until the action is needed. By the time the agent needs to put that learning into effect, the older adult---now experiencing memory problems---may be ready to accept (and expect) support from the agent.

\subsection{Agent's Role Transitions: Tool, Coach, Advisor, and Advocate}

Our findings highlight the need for the agent's role to be dynamic throughout different points of the older adult's experience and care network. CSCW and HCI research generally assume that an agent's role is static. For example, in a study focused on older adults' adoption of stair lifts, older adults viewed the stair lift solely as a tool~\cite{soubutts2021aging}, and did not imagine it to play any other roles at any other times. In another study, Wong et al.~\cite{wong2024voice} explored using conversational agents to support older adults' mental health. The older adults viewed the agent as a companion and did not expect its role to change. In our study, we saw that for older adults, agents may need to transition from being a \textit{tool} to a \textit{coach} to an \textit{advocate}. For caregivers, agents may need to transition from being a \textit{tool} to an \textit{advisor}. There is an opportunity to create agents that learn to change their role to match the needs of the multiple users simultaneously. 

\subsubsection{Tool to Coach to Advocate}
To support older adults, the agent's role change may be gradual: start with being a tool, then evolve to become a coach, and then, later on, an advocate as the older adult's health declines. In addition, the agent's role change may depend on the agent's affiliation, i.e., who it works for, in the older adult's care network~\cite{chang2024dynamic}. There's a lack of understanding of the impacts of agent affiliation on adoption as confirmed by prior work~\cite{luria2020social, chang2024dynamic}.  

\subsubsection{Tool to Advisor}
The agent's role may also need to change over time from the perspective of the caregiver. When caregivers are adult children, agent role change may model the natural role reversal~\cite{mayseless2004more} that occurs in many families as children grow up and their parents age. First, the children are the ones cared for. During this phase, the agent serves as a tool, supporting practical tasks that the parent needs to do, including some that are centered on caring for children.  Later on, the children are the ones providing the care. As this change occurs, the agent becomes an advisor to the adult children, leveraging the deep knowledge of the older adult being supported that it has developed over time and balancing advisory duties with the coaching and advocacy that it provides for the older adult.

\subsubsection{What Does It Mean to be a Coach, Advisor, or Advocate?}
Serving as a coach, advisor, or advocate means understanding factors that influence rapport, trust, and teamwork~\cite{siewiorek2012architecture, robinette2017effect, chang2020tasc}. For instance, agents may need to learn when to keep secrets and when to violate trust to share information that an older adult does not want shared to prioritize the older adult's safety. Prior work shows that people's perception of performance expectancy and intention to hire are the same for robo-advisors and novice financial advisors~\cite{zhang2021you}. There is an opportunity to design interactions that enable agents to be perceived as loyal, trustworthy, and respectful. Aguirre et al.~\cite{aguirre2020ai} argue that a loyal agent will form stronger relationships with users, allowing for more information sharing which will enable the agent to provide a wider range of services, as opposed to services that can be provided only when certain information is shared. In addition, agents should also minimize older adults' fears of being a burden and caregivers' worries about doing a good enough job. 

Open research questions include: What would an agent need to know to become an effective advocate in the care network, and would people listen? Prior work in HRI explored robots as advisors but only in the context of a lab environment, which lacks the real-world challenges of older adults' care networks~\cite{polakow2022social}. Moreover, agents would most likely need to collect large amounts of personal data over long periods of time to serve as an advocate, which may create a new kind of privacy concern. What are the ethical considerations for collecting and analyzing this type of data? Also, how does an agent balance being both an advisor for the caregiver and an advocate for the older adult? For the coach and advisor roles, future work could explore if the agent needs to operate in a space similar to doctors and ministers and follow ethical and legal guidelines set by existent protocols.

\subsection{Unremarkable to Remarkable Agents}
Healthy older adults expressed a preference for an unremarkable agent, that is, an agent that exists in the \textit{background}. They do not want to think about how it may be supporting them. This aligns with prior work showing that older adults desire for fall detector devices to be unremarkable because they are an unwelcome reminder of aging~\cite{caldeira2017senior}. Relating to adoption, health professionals who resist aid from decision support tools are more accepting of AI systems that are unremarkable~\cite{yang2019unremarkable}. Older adults suffering from cognitive decline prefer the agent to be remarkable, existing in the \textit{foreground} and being visible. To the best of our knowledge, this concept of agents transitioning from unremarkable to remarkable has not been explored in prior research.

\subsubsection{Dynamics of Unremarkableness}
This represents an opportunity for the CSCW community to explore this design space of developing agents with dynamic unremarkableness. An agent's unremarkableness may need to change as the agent's role changes such as transitioning from a tool to an advocate. As an advocate, the agent would become remarkable. For agents to achieve an appropriate level of unremarkableness, they will need to model users' daily routines, preferences and relationships over time. Agents will also need to reason about when to insert themselves in the daily routine in a manner that is natural and context-sensitive. For example, agents supporting healthy older adults may need to operate with knowledge that the older adult has mismatched capabilities and reason about when to acknowledge the mismatch. An agent's decision-making may be also dependent on its affiliation~\cite{chang2024dynamic}. If the agent works for the older adult, the agent may not address the mismatched capabilities, but if the agent works for the caregiver, the agent may need to notify that person. This type of reasoning about social situations is lacking in current agent capabilities.

Future work could explore the following research questions: Where does the appropriate level of unremarkableness lie as the older adult's health declines? How does the change occur? What information can help agents learn to make the transition(s)? If agent support is adopted prior to cognitive decline, could the agent serve as a coach to reason with the older adult about what the near future might look like? Our findings show that there is a difference in awareness of health conditions between healthy older adults and those experiencing memory problems, which has been shown in previous research~\cite{forlizzi2004assistive}. In particular, we saw that older adults have a tendency to overestimate their driving skills~\cite{huang2020self}. This may indicate that people's assumptions about their future selves may not be accurate and should not be the only things considered. What other factors should be considered to better capture future selves? When is the appropriate time to ask about one's future self? 

\subsection{Desire for Control of the Agent Shifts Based on Older Adults' Health}
Our findings suggest that participants' desire for control of the agent changes as their health declines. When older adults are healthy, they want to control the agent, but when they experience decline, they want to be cared for. Researchers have studied the tensions between older adults with MCI and their adult children in the context of control of data of in-home monitoring systems. When the older adults lacked understanding of the data being shared, they were hesitant to share the data~\cite{li2023privacy}. This conflicts with our insight that as the older adult's health declines, they want to relinquish control of the agent to their caregivers. This difference in findings may be related to how older adults view the role of the technology. While older adults may be hesitant to share data they do not understand, they might still prefer to delegate control to the agent when they perceive it as a direct caregiver such as their advocate, emphasizing the agent's role and perceived trustworthiness.

\subsubsection{When Might the Control Shift Occur?}
The point of transition for older adults just beginning to experience cognitive decline is likely to be challenging. At this point in an older adult's decline, caregivers also want to control the agent because they feel that they carry all of the responsibility for both preventing breakdowns and maintaining the declining older adult's sense of self. There is an opportunity to design agents that understand the change in power dynamics between older adults and their informal caregivers over time. Prior work shows that aging parents in individualistic, overt-power families maintain power for longer than those in collectivist, hidden-power families, and they experience more conflict and strife during the transition of power from older adult to adult child~\cite{pyke1999micropolitics}. One possibility is to design agents that promote healthy dynamics around aging usually found in collectivist, hidden-power families among members of individualist, overt-power families.

There is also an opportunity to develop agents that manage situations with different conflicts of interest and decide when to assert control on the older adult's behalf. This direction could also build on existing human-robot interaction work on perspective-taking methods~\cite{trafton2005enabling, hiatt2004cognitive} to gain insight into when and how to assert control. For instance, agents take on the older adult's perspective when making a decision for an older adult whose decision-making power is diminished: they could draw on data about how the older adult used to make decisions and come up with a plan that aligns as closely as possible with what the older adult likely would have done. 

As the older adult declines, they cede power and agency to their caregivers. If the agent absorbs some of the burden of care, an interesting research question is whether or not the agent should also receive some of that agency. Another open research question is how an agent should make decisions about multiple, competing objectives in the case where the older adult wants the agent to assert control on their behalf and the caregiver wants to control the agent.

\subsubsection{Ethical Considerations}
There are many design opportunities to advance the development of agent support for older adults as they continue to age at home. However, there is potential for unintended harm and other negative outcomes. There seems to be an opportunity to rethink ways to minimize unintended harm and understand the true benefits of agent support, especially for this target population. This research space comes with additional ethical complexities: some users are in denial of their cognitive decline, and most situations are likely to have conflicts of interests among the various stakeholders in the care network. AI ethical frameworks such as the one from London \& Heidari~\cite{london2023beneficent} can be leveraged to assess agents' potential decisions to avoid behavior related to domination, deception, exploitation, and unjustified paternalism.

\section{Limitations}
Our work has several limitations due to its exploratory nature. For example, storyboards do not involve interacting with an actual agent---people’s actual acceptance and behaviors may not be accurate. Long-term deployment of real systems is needed to understand real behavior. We view our findings and insights as nascent theory, regional to the U.S. and not global. The distinctions we draw between feedback from healthy older adults and those with memory impairments might be misattributed to population differences when there are actually other variables at play. Also, there are many other stakeholders whose perspectives we did not seek (e.g., other relatives, clinicians, professional caregivers), but which may influence the feasibility of our design considerations.

\section{Conclusion}
We used speed dating to explore the boundaries of agent acceptance to providing care support for older adults as they navigate through different life stages. We discovered that boundaries of acceptable agent behavior in the older adult's care network changes according to the older adult's cognitive decline and emotions associated with that decline. That is, initially, participants only want the agent to provide functional support, but when they experience decline, they want the agent to provide social support such as advising and advocating. This suggests that agents may need to evolve from being unremarkable to remarkable. This builds on prior research that views adoption as related to only functional support. We discovered that healthy older adults are hesitant to adopt agent support. However, when they experience cognitive decline, they want support from an agent to be personalized as if they have adopted it when they were healthy. In contrast, older adults experiencing memory problems and their caregivers are willing to use the agent now and wish to have adopted it well before the start of cognitive decline. One of the barriers to adoption for healthy older adults is their fear of cognitive impairment and their sense that accepting an agent is surrendering to this possibility. This suggests a need for researchers, designers, and developers to consider adoption at a much earlier life stage.

\begin{acks}
We thank the members and staff of the senior center where we conducted our research, our colleagues across the NSF AI-CARING Institute for their feedback, and REU students (Irene Kang and Winnie Lin) and pilot participants for their work on the storyboards. This research is supported by the National Science Foundation (IIS-2112633).
\end{acks}

\bibliographystyle{ACM-Reference-Format}
\bibliography{refs}

\appendix

\section{Appendix}
\subsection{Storyboards}
The storyboards for the \href{https://tinyurl.com/cscw200-storyboards-healthy}{healthy group of participants}. 
\newline
The storyboards for the \href{https://tinyurl.com/cscw200-storyboards-declining}{declining group of participants}.
\newline
The \href{https://tinyurl.com/cscw200-questions}{leading questions} for each storyboard.

\end{document}